\documentclass[twocolumn,showpacs,preprintnumbers]{revtex4-2}
\usepackage[caption=false]{subfig}
\usepackage{amsmath}  
\usepackage{amsfonts}  
\usepackage{amssymb}
\usepackage{graphicx} 
\usepackage{enumerate}
\usepackage{tikz}
\usepackage{soul}
\usetikzlibrary{shapes,arrows, arrows.meta, calc, positioning}
\usepackage[colorlinks=true,linkcolor=blue,urlcolor=blue,citecolor=blue]{hyperref}
\usepackage{xcolor}
\usepackage[normalem]{ulem} 
\begin{document}

\newcommand{\ds}{\displaystyle}
\newcommand{\be}{\begin{equation}}
\newcommand{\en}{\end{equation}}
\newcommand{\bea}{\begin{eqnarray}}
\newcommand{\ena}{\end{eqnarray}}

\title{Geometric origin of the cosmological constant from Einstein-Chern-Simons gravity compactified to four dimensions}

\author{M.~Cataldo}
\email{mcataldo@ubiobio.cl}
\affiliation{Departamento de F\'{\i}sica, Universidad del B\'{\i}o-B\'{\i}o, Casilla 5-C, Concepci\'{o}n, Chile.}
\affiliation{Centro de Ciencias Exactas, Universidad del B\'{\i}o-B\'{\i}o, Casilla 447, Chill\'{a}n, Chile.}

\author{S.~Lepe}
\email{samuel.lepe@pucv.cl}
\affiliation{Instituto de F\'{\i}sica, Pontificia Universidad Cat\'{o}lica de Valpara\'{\i}so, Avda.\ Brasil 2950, Valpara\'{\i}so, Chile.}

\author{C.~Riquelme}
\email{ceriquelme@udec.cl}
\affiliation{Departamento de F\'{\i}sica, Universidad de Concepci\'{o}n, Casilla 160-C, Concepci\'{o}n, Chile.}

\author{P.~Salgado}
\email{patsalgado@unap.cl}
\affiliation{Facultad de Ciencias, Universidad Arturo Prat, Avda.\ Arturo Prat 2120, Iquique, Chile.}
\affiliation{Instituto de Ciencias Exactas y Naturales, Avda.\ Playa Brava 3256, Iquique, Chile.}

\date{\today}

\begin{abstract}
We present a model in which the cosmological constant emerges as a purely geometric effect from the four-dimensional compactification of five-dimensional Einstein--Chern--Simons gravity. Within this framework, the compactification of the extra dimension gives rise to an effective cosmological constant $\Lambda$ that depends on the compactification radius $r_c$, the dimensional coupling parameter $l$, and the trace $\tilde{h}$ of the compactified field $h^a$, rather than being introduced as an ad hoc free parameter. We demonstrate that the resulting field equations are structurally equivalent to those of General Relativity with a cosmological constant. Consequently, all known vacuum solutions---including the Schwarzschild--de Sitter, Kerr--de Sitter, and FLRW spacetimes---remain valid in this setting. As a concrete application, we explicitly derive the Kottler (Schwarzschild--de Sitter) black hole solution.

We identify two distinct dynamical regimes. In the weak-field regime, $\Lambda \propto l^{2}\tilde{h}/r_{c}^{3}$ and its sign is controlled by $l^2\tilde{h}$, requiring fine-tuning of the coupling constants to reproduce the observed value $\Lambda_{\text{obs}} \approx 10^{-52}\,\text{m}^{-2}$. In the strong-field regime, the dependence on $l$ and $\tilde{h}$ drops out through algebraic cancellation and one obtains $\Lambda \approx 3/(4r_{c}^{2})$, independently of the Chern-Simons coupling. This regime naturally reproduces $\Lambda_{\rm obs}$ for a compactification radius $r_{c} \approx 0.78\,H_{0}^{-1} \approx 8.2 \times 10^{25}\,\text{m}$, without any fine-tuning, with the cosmological constant determined purely by the compactification radius, and the Bekenstein-Hawking entropy of the cosmological horizon yields $S_{\rm cosm} = 4\pi k_B r_c^2/l_{\rm Pl}^2 \sim 10^{122}\,k_B$, in agreement with the Gibbons-Hawking result and admitting a direct geometric interpretation in terms of $r_c$.

This framework reframes the cosmological constant problem geometrically: rather than asking why $\Lambda$ is so small, one asks why the compactification radius is so large---a reformulation consistent with the possibility of coexisting with a large extra dimension without violating established gravitational tests.
\end{abstract}

\maketitle

\section{Introduction}

The formulation of gravity as a gauge theory requires, as a fundamental principle, that the action be constructed independently of any fixed spacetime background. In this sense, the dynamical geometry itself must emerge from the gauge structure, rather than being imposed \emph{a priori}. A natural realization of this idea is provided by Chern--Simons (ChS) theories, whose actions are invariant under a given gauge algebra and do not rely on a background metric.

In particular, a gravitational Lagrangian satisfying these requirements can be formulated as a Chern--Simons form for the (Anti-)de Sitter algebra. Such a construction leads to a geometrically motivated action principle in which the vielbein $e^{a}$ and the spin connection $\omega^{ab}$ arise as components of a single gauge connection valued in the (A)dS algebra~\cite{champ1,champ2,zan1}.

In five dimensions, the corresponding Chern--Simons action can be written explicitly as
\begin{align}
S_{(\mathrm{A})\mathrm{dS}}^{\left( 5\right) } 
&= \frac{1}{8\kappa_5} 
\int  \pm \frac{1}{5l^{2}}\varepsilon_{abcde}\,e^{a}e^{b}e^{c}e^{d}e^{e}
\nonumber\\
&\quad + \frac{2}{3}\varepsilon_{abcde}\,R^{ab}e^{c}e^{d}e^{e}
\pm l^{2}\varepsilon_{abcde}\,R^{ab}R^{cd}e^{e},
\label{1}
\end{align}
which is off-shell invariant under the AdS Lie algebra $\mathfrak{so}(4,2)$ and dS Lie algebra $\mathfrak{so}(5,1)$ respectively. Here, $\kappa_5$ is the five-dimensional gravitational coupling constant, $\varepsilon_{abcde}$ is the five-dimensional Levi-Civita symbol, $e^{a}$ is the 1-form vielbein, and $R^{ab} = \mathrm{d}\omega^{ab} + \omega^{a}{}_c \omega^{cb}$ 
is the curvature 2-form in the first-order formalism, obtained from the 1-form spin connection $\omega^{ab}$, which in turn is related to the Riemann curvature tensor. The coupling constant $l$, which has dimensions of length, characterizes different regimes within the theory, and the $\pm$ signs reflect the two possible branches of the theory, whose physical interpretation is determined by the compactification procedure, as shown in Sec.~\ref{sec:einstein}.

From action (\ref{1}), it is apparent that neither $l\rightarrow \infty $ the nor the $l\rightarrow 0$ limit yields the Einstein-Hilbert term $\varepsilon _{a_{1}\cdots a_{5}}R^{a_{1}a_{2}}e^{a_{3}}\cdots e^{a_{5}}$ on its own. Such limits lead either to the Gauss-Bonnet term or to the cosmological constant term, respectively.

If Chern-Simons (ChS) theories are the appropriate gauge theories to provide a framework for gravitational interaction, then they must satisfy the correspondence principle; namely, they must be reducible to General Relativity (GR).

In a previous work, it was shown that standard five-dimensional (5D) GR can be obtained from a ChS gravity theory based on a certain Lie algebra $\mathfrak{B}_{5}$, whose generators satisfy the commutation relations shown in Eq. (7) of Ref. \cite{gomez}. This algebra was obtained from the Anti-de Sitter (AdS) algebra and a particular semigroup $S$ by means of the S-expansion procedure introduced in Refs. \cite{salg2}, \cite{salg3}, \cite{azcarr}.

It is straightforward to prove that GR in 5D can also be obtained from a ChS gravity theory based on a Lie algebra $\mathfrak{\tilde{B}}_{5}$, which is derived from the de Sitter (dS) algebra using the aforementioned S-expansion method. The only difference between $\mathfrak{B}_{5}$ and $\mathfrak{\tilde{B}}_{5}$  lies in the sign of the commutator between generalized
translations $\left[ \boldsymbol{P}_{a},\boldsymbol{P}_{b}\right] $ from the algebra shown in Eq. (7) of Ref. \cite{gomez}.

By applying Theorem VII.2 of Ref. \cite{salg2} and the extended Cartan's homotopy formula as in Ref.  \cite{salg4}, it was found that the five-dimensional ChS Lagrangian for the $\mathfrak{B}_{5}$ and $\mathfrak{\tilde{B}}_{5}$ algebras is given by 
\begin{align}
L_{\text{EChS}}^{(5)} &= \alpha_{1}\left(\pm l^{2}\right) 
\varepsilon_{abcde}e^{a}R^{bc}R^{de} \notag \\
& \quad +\alpha_{3}\varepsilon_{abcde}\left(\frac{2}{3}
R^{ab}e^{c}e^{d}e^{e}\right. \notag \\
& \quad \left. +\left(\pm l^{2}\right) R^{ab}R^{cd}h^{e}
+2\left(\pm l^{2}\right) k^{ab}R^{cd}T^{e}\right), \label{eq:lagrangian}
\end{align}
where $\alpha _{1}$ and $\alpha _{3}$ are parameters of the theory, $l$ is a coupling constant, $R^{ab}$ is the curvature 2-form previously introduced,\ $e^{a}$ is the vielbein, and $h^{a}$ and $k^{ab}$ are other gauge fields present in the theory \cite{salg1}.

The field content induced by the $\mathfrak{B}_{5}$ algebra includes the vielbein $e^{a}$, the spin connection $\omega^{ab}$, and two extra bosonic fields, $h^{a}$ and $k^{ab}$. From Eq.~\eqref{eq:lagrangian}, it can be seen that the five-dimensional Einstein gravity can be recovered from Chern-Simons gravity in the limit where the coupling constant $l$ vanishes while keeping the effective Newton's constant fixed. It was proven in Ref.~\cite{quin} that the field equations obtained by varying action~\eqref{eq:lagrangian} admit black hole solutions (see also Ref.~\cite{crisos}).

Furthermore, in Refs.~\cite{gomez1,PS 2021}, it was found that five-dimensional Einstein-Chern-Simons gravity is consistent with a four-dimensional spacetime. In fact, these references established that the Randall-Sundrum~\cite{randall,randall1} compactification of action~\eqref{eq:lagrangian} leads to the following four-dimensional action:
\begin{align}
\tilde{S}[\tilde{e},\tilde{h}] &= \int_{\Sigma_{4}} \tilde{\varepsilon}_{mnpq} \left( -\frac{1}{2} \tilde{R}^{mn} \tilde{e}^{p} \tilde{e}^{q} \right. \notag \\
&\quad \left. + K \tilde{R}^{mn} \tilde{e}^{p} \tilde{h}^{q} - \frac{K}{4r_{c}^{2}} \tilde{e}^{m} \tilde{e}^{n} \tilde{e}^{p} \tilde{h}^{q} \right),
\label{eq:action4D}
\end{align}
where $\tilde{\varepsilon}_{mnpq}$, $\tilde{e}^{m}$, $\tilde{R}^{mn}$, and $\tilde{h}^{m}$ represent, respectively, the $4$-dimensional versions of the Levi-Civita symbol, the vielbein, the curvature form, and a matter field, while $r_{c}$ is the so-called compactification radius. The parameter $K$ is a constant defined as~\cite{PS 2021}
\begin{equation}
K = \frac{\pm l^{2} \pi}{2 \kappa r_{c}}, \label{KK}
\end{equation}
where $\kappa$ is the four-dimensional gravitational coupling constant. It is important to emphasize that the field $h^{a}$ induces a cosmological constant arising from the compactification of the five-dimensional Einstein--Chern--Simons theory to four dimensions. Its sign, positive (de Sitter) or negative (Anti-de Sitter), is fixed by the sign of the product $l^{2}\tilde{h}$, as demonstrated explicitly in Sec.~\ref{sec:einstein}.

The origin of the cosmological constant remains one of the most profound open problems in theoretical physics. Quantum field theory estimates for the vacuum energy density exceed the observed value by approximately 120 orders of magnitude--a discrepancy often regarded as the worst prediction in the history of physics~\cite{weinberg,Weinberg1,Adler}. In standard General Relativity, $\Lambda$ must be introduced as a free parameter without any fundamental explanation for its small but nonzero value. The framework presented here offers a conceptually different perspective: the cosmological constant emerges naturally from the compactification of Einstein-Chern-Simons gravity, with its magnitude determined by the compactification radius $r_{c}$ rather than being put in by hand. As we shall demonstrate, the observed value $\Lambda_{\text{obs}} \approx 10^{-52}\,\text{m}^{-2}$ corresponds to $r_{c} \sim 10^{26}\,\text{m}$, remarkably close to the current Hubble radius. While this does not solve the cosmological constant problem in the fundamental sense--since one must still explain why $r_{c}$ takes this particular value--it provides a geometric mechanism that relates $\Lambda$ to the large-scale structure of spacetime, potentially offering new avenues for understanding dark energy.

This article is organized as follows. In Sec.~\ref{sec:einstein}, we show that the field equations obtained from the four-dimensional compactification of five-dimensional Einstein-Chern-Simons gravity coincide with the Einstein field equations with a cosmological constant that depends on the trace of the compactified field and the compactification radius $r_{c}$. In Sec.~\ref{sec:spherical}, we derive the explicit field equations for a static, spherically symmetric metric and show that the most general solution is a Kottler-type (Schwarzschild-de Sitter) black hole. In Sec.~\ref{sec:cosmology}, we analyze the implications of this framework for $\Lambda$CDM cosmology: we identify two distinct dynamical regimes of the cosmological constant, derive an estimate for the compactification radius, discuss the embedding of $\Lambda$CDM within Einstein-Chern-Simons gravity, and examine the geometric reformulation of the cosmological constant problem. Finally, Sec.~\ref{sec:conclusions} summarizes our main results and outlines directions for future work.

\section{Einstein Field Equations from Einstein-Chern-Simons Gravity}
\label{sec:einstein}

In this section, we show that the field equations~\eqref{eq:fieldG} and~\eqref{eq:fieldV} coincide with the Einstein field equations with a specific cosmological constant, which depends on the trace $\tilde{h}$ of the four-dimensional component of the $5D$ field $h^{a}$ and the compactification radius $r_{c}$. Indeed, in tensorial language, the action~\eqref{eq:action4D} takes the form~\cite{gomez1,PS 2021}
\begin{equation}
\tilde{S}[\tilde{g},\tilde{h}] = \int d^{4}\tilde{x} \sqrt{-\tilde{g}} \left[ \tilde{R} + 2K \left( \tilde{R}\tilde{h} - 2\tilde{R}^{\mu}_{\ \nu} \tilde{h}^{\nu}_{\ \mu} \right) - \frac{3K}{2r_{c}^{2}} \tilde{h} \right],
\label{eq:actiontensor}
\end{equation}
where it can be seen that as $l \longrightarrow 0$ then $K \longrightarrow 0$, and actions~\eqref{eq:action4D} and~\eqref{eq:actiontensor} become the four-dimensional Einstein-Hilbert action. The field $\tilde{h}_{\mu\nu}$, inherited from the five-dimensional theory, is in principle an independent symmetric tensor. However, since we seek solutions that reduce to standard General Relativity with a cosmological constant---thus preserving all classical tests of gravity---the natural requirement is that $\tilde{h}_{\mu\nu}$ contributes to the field equations only through a term proportional to the metric tensor, mimicking the structure of a cosmological term $\Lambda g_{\mu\nu}$. This is achieved by the Ansatz
\begin{equation}
\tilde{h}_{\mu\nu} = \frac{1}{4} \tilde{F}(\tilde{\phi}) \tilde{g}_{\mu\nu},
\label{eq:ansatz}
\end{equation}
where $\tilde{F}$ is a function of a scalar field $\tilde{\phi} = \tilde{\phi}(\tilde{x}^{\mu})$, with $\tilde{x}^{\mu} = (x^{0}, x^{i})$ for $i = 1, 2, 3$. This choice guarantees that the resulting field equations retain the form of Einstein's equations with an effective cosmological constant, ensuring that all known solutions of General Relativity with $\Lambda$--such as Schwarzschild-de Sitter, Kerr-de Sitter, and FLRW spacetimes--remain valid within this framework. The physical content of the compactification is then encoded entirely in the geometric origin of $\Lambda$, rather than in modifications to the structure of the solutions themselves. With this Ansatz, we can write
\begin{equation}
\tilde{R}^{\mu}_{\ \nu} \tilde{h}^{\nu}_{\ \mu} = \frac{1}{4} \tilde{F}(\tilde{\phi}) \tilde{R}, \qquad \tilde{h} = \tilde{h}_{\mu\nu} \tilde{g}^{\mu\nu} = \tilde{F}(\tilde{\phi}),
\end{equation}
such that action~\eqref{eq:actiontensor} takes the form
\begin{equation}
\tilde{S}[\tilde{g},\tilde{\phi}] = \int d^{4}\tilde{x} \sqrt{-\tilde{g}} \left[ \tilde{R} + K\tilde{R}\tilde{h}(\tilde{\phi}) - \frac{3K}{2r_{c}^{2}} \tilde{h}(\tilde{\phi}) \right],
\label{eq:actionscalar}
\end{equation}
which corresponds to an action for four-dimensional gravity non-minimally coupled to a scalar field. Note that this action takes the form $\tilde{S} = \tilde{S}_{g} + \tilde{S}_{g\phi} + \tilde{S}_{\phi}$, where $\tilde{S}_{g}$ is the pure gravitational term, $\tilde{S}_{g\phi}$ is the non-minimal interaction term between gravity and the scalar field, and $\tilde{S}_{\phi}$ represents a scalar field potential term. In order to write the action in a more suitable form, we define the constant $\varepsilon$ and the potential $V(\tilde{\phi})$ as
\begin{equation}
\varepsilon = \frac{4\kappa r_{c}^{2}}{3}, \qquad V(\tilde{\phi}) = \frac{3K}{4\kappa r_{c}^{2}} \tilde{h}(\tilde{\phi}).
\label{eq:definitions}
\end{equation}
This allows us to rewrite action~\eqref{eq:actionscalar} as
\begin{equation}
S[g,\phi] = \int d^{4}x \sqrt{-g} \left[ \tilde{R} + \varepsilon \tilde{R} V(\tilde{\phi}) - 2\kappa V(\tilde{\phi}) \right].
\label{eq:actionfinal}
\end{equation}
Varying this action yields the following field equations~\cite{gomez1,PS 2021}:
\begin{equation}
G_{\mu\nu} = -\kappa \left( \frac{V}{1 + \varepsilon V} \right) g_{\mu\nu},
\label{eq:fieldG}
\end{equation}
\begin{equation}
\frac{\partial V}{\partial \tilde{\phi}} \left( 1 - \frac{\varepsilon \tilde{R}}{2\kappa} \right) = 0.
\label{eq:fieldV}
\end{equation}
Using the Bianchi identity $\nabla^{\mu} G_{\mu\nu} = 0$ and the metricity condition $\nabla_{\rho} g_{\mu\nu} = 0$, we find that
\begin{equation}
\nabla_{\mu} \left( \frac{V}{1 + \varepsilon V} \right) = \partial_{\mu} \left( \frac{V}{1 + \varepsilon V} \right) = 0,
\label{eq:bianchi}
\end{equation}
implying that $V(\tilde{\phi})$ is constant. This means that Eq.~\eqref{eq:fieldG} coincides with the Einstein field equations with a cosmological constant $\Lambda$ that depends on the compactification radius $r_{c}$, the coupling constant $l$, and the scalar field $\tilde{h}$. This implies that the cosmological constant in question may be interpreted as a consequence of $5D$ to $4D$ compactification. It is important to note that Eq.~\eqref{eq:fieldG} can be rewritten in the form
\begin{equation}
G_{\mu\nu} = \kappa T_{\mu\nu}^{\text{vac}},
\end{equation}
where $T_{\mu\nu}^{\text{vac}} = -\rho_{\text{vac}} g_{\mu\nu}$ represents the energy-momentum tensor of the vacuum with energy density $\rho_{\text{vac}} = V/(1 + \varepsilon V)$. This structure places the cosmological term on the right-hand side of Einstein's equations, suggesting an interpretation in terms of vacuum energy rather than intrinsic spacetime curvature. However, defining $\Lambda \equiv \kappa \rho_{\text{vac}}$, the field equations become
\begin{equation}
G_{\mu\nu} + \Lambda g_{\mu\nu} = 0, \label{EEQ}
\end{equation}
which is the standard form with $\Lambda$ as a purely geometric quantity. Thus, the two interpretations---the cosmological constant as vacuum energy density versus intrinsic curvature of spacetime---are mathematically equivalent~\cite{Carroll}. The distinction is conceptual: in the present framework, $\Lambda$ emerges from the compactification of the fifth dimension, providing a geometric origin that transcends both interpretations.

Indeed, using Eq.~\eqref{eq:fieldG} and~\eqref{EEQ} it is straightforward to show that the cosmological constant takes the explicit form
\begin{eqnarray}
\Lambda = \kappa  \, \frac{V}{1 + \varepsilon V}. \label{eq:lambda}
\end{eqnarray}
The sign structure of the cosmological constant~\eqref{eq:lambda} 
can be systematically analyzed by examining the signs of $V$ and 
$1 + \varepsilon V$ independently. This leads to three physically distinct branches, 
summarized in Table~\ref{tab:branches}.
\begin{table}[h]
\centering
\begin{tabular}{ccccc}
\hline\hline
Branch & $\mathrm{sgn}(V)$ & $\mathrm{sgn}(1+\varepsilon V)$ 
& Condition & Spacetime \\
\hline
1 & $V > 0$ & $1 + \varepsilon V > 0$ & $V > 0$ 
& dS, $0 < \Lambda < \kappa/\varepsilon$ \\
2 & $V < 0$ & $1 + \varepsilon V < 0$ & $V < -1/\varepsilon$ 
& dS, $\Lambda > \kappa/\varepsilon$ \\
3 & $V < 0$ & $1 + \varepsilon V > 0$ & $-1/\varepsilon < V < 0$ 
& AdS, $\Lambda < 0$ \\
\hline\hline
\end{tabular}
\caption{Sign analysis of $\Lambda = \kappa V/(1+\varepsilon V)$ 
from Eq.~\eqref{eq:lambda}. The fourth combination ($V > 0$ with $1 + \varepsilon V < 0$) is algebraically excluded, given that $\varepsilon > 0$. The value $\Lambda = \kappa/\varepsilon = 
3/(4r_c^2)$ is an asymptote approached from below in Branch~1 
and from above in Branch~2, as discussed in Sec.~\ref{sec:cosmology}.}
\label{tab:branches}
\end{table}
In Branch~1, the upper bound $\Lambda < \kappa/\varepsilon$ follows 
directly from Eq.~\eqref{eq:lambda}. Setting $u = \varepsilon V > 0$, 
one obtains $\Lambda = (\kappa/\varepsilon)\, u/(1+u)$, where 
$0 < u/(1+u) < 1$ for all finite $u > 0$. Thus $\Lambda$ is strictly 
bounded above by $\kappa/\varepsilon$, a value that is 
approached asymptotically as $V \to +\infty$ but never reached. 
Analogously, in Branch~2, setting $u = \varepsilon V < -1$ one finds 
$u/(1+u) > 1$, so that $\Lambda > \kappa/\varepsilon$, with the same 
asymptote approached from above as $V \to -\infty$.

By considering Eqs.~\eqref{KK} and~\eqref{eq:definitions} we have that 
\begin{eqnarray*}
V =\pm \frac{3\pi l^2 \tilde{h}}{8\kappa^2 r_{c}^{3}} 
\end{eqnarray*}
and equivalently we may write for the cosmological constant
\begin{align}
\Lambda = \kappa \frac{3 \pi \left(\pm l^{2} \tilde{h} \right) }{3 \pi \varepsilon \left(\pm l^{2}\tilde{h} \right) +8 \kappa^2 r_{c}^{3}} 
= \frac{3 \pi \left(\pm l^{2} \tilde{h} \right) }{4 \pi r_{c}^{2} \left(\pm l^{2} \tilde{h} \right)  +8 \kappa r_{c}^{3}}.\label{eq:lambda_equivalente}
\end{align}
In this expression, the $\pm$ signs reflect the fact that the product $l^2 \tilde{h}$ may be negative or positive, leading to either a negative cosmological constant (Anti-de Sitter, $\Lambda < 0$) or a positive one (de Sitter, $\Lambda > 0$). Consistency with the observed accelerated expansion of the universe requires the positive branch.

Several important features of this result deserve emphasis:

\textit{(i) Geometric origin of $\Lambda$:} The cosmological constant emerges naturally from the compactification procedure, without being introduced by hand into the action. This provides a geometric explanation for dark energy within the framework of Einstein-Chern-Simons gravity.

\textit{(ii) Dependence on the coupling constant:} In the limit $l \rightarrow 0$, we have $\Lambda \rightarrow 0$, recovering standard General Relativity without a cosmological constant. This shows that $\Lambda$ is intrinsically linked to the Chern-Simons corrections and would vanish in pure Einstein gravity.

It is worth emphasizing that the field equations~\eqref{eq:fieldG} are structurally identical to the Einstein field equations with a cosmological constant. Consequently, all known vacuum solutions of General Relativity with $\Lambda$--including Schwarzschild-de Sitter, Kerr-de Sitter, and FLRW spacetimes--remain valid within this framework. The key difference lies not in the geometric structure of the solutions, but in the origin and interpretation of $\Lambda$: rather than being a free parameter, it is now determined by the compactification radius $r_c$ and the coupling constants of the five-dimensional Einstein-Chern-Simons theory, as expressed in Eq.~\eqref{eq:lambda}.

\section{Field Equations for a Spherically Symmetric Metric}
\label{sec:spherical}
In this section, we derive the explicit field equations for a static, spherically symmetric spacetime. We begin by computing the relevant geometric quantities---the Ricci tensor components, the scalar curvature, and the Einstein tensor---for a general metric of this form. These expressions are then substituted into the field equations~\eqref{eq:fieldG} and~\eqref{eq:fieldV} to obtain a system of ordinary differential equations, which we subsequently solve to find the most general static solution.

In four dimensions, a static and spherically symmetric metric can be written in the form:
\begin{equation}
ds^{2} = -e^{\alpha(r)} dt^{2} + e^{\beta(r)} dr^{2} + r^{2} \left( d\theta^{2} + \sin^{2}\theta \, d\varphi^{2} \right),
\label{eq:metric}
\end{equation}
where $r$, $\theta$, and $\varphi$ are the usual spherical polar coordinates. For this metric, the non-vanishing components of the Ricci tensor, the scalar curvature, and the Einstein tensor are given by
\begin{align}
R_{00} &= \frac{e^{\alpha-\beta}}{2} \left( \alpha'' + \frac{\alpha'^{2}}{2} - \frac{\alpha'\beta'}{2} + \frac{2\alpha'}{r} \right), \notag \\
R_{11} &= -\frac{1}{2} \left( \alpha'' - \frac{\alpha'\beta'}{2} - \frac{2\beta'}{r} + \frac{\alpha'^{2}}{2} \right), \notag \\
R_{22} &= 1 - e^{-\beta} \left( 1 + \frac{r}{2}(\alpha' - \beta') \right), \notag \\
R_{33} &= R_{22} \sin^{2}\theta, \notag \\
R &= \frac{2}{r^{2}} - e^{-\beta} \left( \alpha'' + \frac{\alpha'^{2}}{2} - \frac{\alpha'\beta'}{2} + \frac{2}{r}(\alpha' - \beta') + \frac{2}{r^{2}} \right),
\label{eq:ricci}
\end{align}
and
\begin{align}
G_{00} &= \frac{e^{\alpha}}{r^{2}} + e^{\alpha-\beta} \left( \frac{\beta'}{r} - \frac{1}{r^{2}} \right), \notag \\
G_{11} &= \frac{\alpha'}{r} + \frac{1}{r^{2}} - \frac{e^{\beta}}{r^{2}}, \notag \\
G_{22} &= -\frac{e^{-\beta}}{2} \left( \frac{\alpha'\beta'}{2} r^{2} - \frac{\alpha'^{2}}{2} r^{2} - \alpha'' r^{2} - r(\alpha' - \beta') \right), \notag \\
G_{33} &= G_{22} \sin^{2}\theta,
\label{eq:einstein}
\end{align}
where the prime denotes a derivative with respect to $r$. Substituting Eqs.~\eqref{eq:metric} and~\eqref{eq:einstein} into Eq.~\eqref{eq:fieldG}, we obtain the following field equations:
\begin{align}
\left[ \frac{e^{\alpha}}{r^{2}} + e^{\alpha-\beta} \left( \frac{\beta'}{r} - \frac{1}{r^{2}} \right) \right] (1 + \varepsilon V) &= \kappa e^{\alpha} V, \label{eq:uno} \\
\left[ \frac{\alpha'}{r} + \frac{1}{r^{2}} - \frac{e^{\beta}}{r^{2}} \right] (1 + \varepsilon V) &= -\kappa e^{\beta} V, \label{eq:dos} \\
\frac{1}{4} \left[ 2\alpha'' r - (\beta' - \alpha')(\alpha' r + 2) \right] (1 + \varepsilon V) &= -\kappa r e^{\beta} V. \label{eq:tres}
\end{align}

Thus far, we have only considered Eq.~\eqref{eq:fieldG}, but have not yet employed Eq.~\eqref{eq:fieldV}, which can be written in the form
\begin{equation}
\frac{\partial V}{\partial r} \left( 1 - \frac{\varepsilon R}{2\kappa} \right) = 0.
\label{eq:z4}
\end{equation}
Indeed, since $\tilde{\phi} = \tilde{\phi}(\tilde{x}^{\mu})$, spherical symmetry requires that $\tilde{\phi} = \tilde{\phi}(r)$, so that Eq.~\eqref{eq:fieldV} reduces to Eq.~\eqref{eq:z4}.

From Eq.~\eqref{eq:uno}, we can obtain the function $V(r)$, which, when substituted into Eq.~\eqref{eq:dos}, yields
\begin{equation}
\frac{\kappa r e^{\beta(r)} \left[ \alpha'(r) + \beta'(r) \right]}{-\varepsilon r \beta'(r) + (r^{2}\kappa - \varepsilon) e^{\beta(r)} + \varepsilon} = 0.
\label{eq:z5}
\end{equation}
This differential equation yields
\begin{equation}
\alpha(r) = c_{1} - \beta(r).
\label{eq:z6}
\end{equation}
Substituting this result, along with the expression for $V(r)$ obtained from Eq.~\eqref{eq:uno}, into Eq.~\eqref{eq:tres} yields the following differential equation for the function $\beta(r)$:
\begin{equation}
\frac{\left[ r^{2} \beta'(r)^{2} - r^{2} \beta''(r) - 2 + 2e^{\beta(r)} \right] r^{2}\kappa}{-2\varepsilon r \beta'(r) + (2r^{2}\kappa - 2\varepsilon) e^{\beta(r)} + 2\varepsilon} = 0,
\label{eq:z7}
\end{equation}
from which we obtain
\begin{equation}
\beta(r) = -\ln\left( 1 - \frac{c_{1}}{r} + \frac{c_{2}}{3} r^{2} \right) = c_{1} - \alpha(r),
\label{eq:z8}
\end{equation}
which is the most general solution to the system of Eqs.~\eqref{eq:uno}--\eqref{eq:tres}, and also identically satisfies Eq.~\eqref{eq:z4}.

Substituting this result into the expression for $V(r)$ obtained from Eq.~\eqref{eq:uno}, we find that:
\begin{equation}
V(r) = -\frac{c_{2}}{\varepsilon c_{2} + \kappa},
\label{eq:z9}
\end{equation}
that is, $V$ is a constant, confirming what was established previously.

The line element corresponding to Eq.~\eqref{eq:z8} depends on the sign of $c_{1}$. Thus, the function $\beta(r)$ takes different forms depending on whether $c_{1}$ is positive or negative. If $c_{1} = 0$, we obtain the usual solution that leads to a black hole with a cosmological event horizon. For $c_{1} = r_{s} = 2M > 0$ and $c_{2} = -\Lambda$, we obtain the Kottler black hole, 
\begin{align}
ds^{2} &= -\left( 1 - \frac{r_{s}}{r} - \frac{\Lambda}{3} r^{2} \right) dt^{2} \notag \\
&\quad + \left( 1 - \frac{r_{s}}{r} - \frac{\Lambda}{3} r^{2} \right)^{-1} dr^{2} + r^{2} d\Omega^{2},
\label{eq:kottler}
\end{align}
also known as the Schwarzschild-de Sitter black hole, where $r_{s}$ is the Schwarzschild radius and $\Lambda$ is the cosmological constant, which in our case is given by Eq.~\eqref{eq:lambda}.

In Ref.~\cite{randall1}, Randall and Sundrum established that it is possible to consistently coexist with a large, or even infinite, fifth dimension without violating known tests of gravity. While it is generally accepted that any extra dimension playing a role in our universe would be difficult to detect due to its small size (a view supported by both theoretical and experimental arguments), the framework presented here suggests an alternative: an extra dimension with a compactification radius $r_{c}$ of cosmological scale, whose presence manifests not through deviations from Newtonian gravity, but through the emergence of the cosmological constant $\Lambda$ via the relation established in Eq.~\eqref{eq:lambda}.

\section{$\Lambda$CDM Cosmology from Einstein-Chern-Simons Compactification}
\label{sec:cosmology}

The $\Lambda$CDM model constitutes the standard framework of modern cosmology, successfully describing the large-scale structure and evolution of the universe in terms of a cosmological constant $\Lambda$, cold dark matter, baryonic matter, and radiation. Observational evidence from Type Ia supernovae, the cosmic microwave background (CMB), and baryon acoustic oscillations (BAO) strongly supports this model, indicating that approximately 68\% of the energy content of the universe consists of dark energy, well characterized by $\Lambda \approx 10^{-52}\,\text{m}^{-2}$. Despite its remarkable observational success, the model faces growing tensions between early- and late-universe probes, including a statistically significant discrepancy in the determination of the Hubble constant $H_0$, hints of dynamical dark energy with equation of state $w \neq -1$, and a mild but persistent discrepancy in the amplitude of matter fluctuations. These anomalies suggest that $\Lambda$CDM, while an excellent approximation to reality, may be incomplete, and motivate the search for theoretical frameworks that provide a deeper origin for the cosmological constant.
However, within standard General Relativity, the cosmological constant must be introduced as a free parameter with no fundamental explanation for its value. This is particularly troubling given that quantum field theory estimates of the vacuum energy density exceed the observed value by some 121 orders of magnitude---a discrepancy known as the cosmological constant problem. The $\Lambda$CDM model, while phenomenologically successful, simply adopts $\Lambda$ as an empirical input and offers no resolution to this puzzle.

The framework developed in this work has a natural and significant implication for $\Lambda$CDM cosmology. As demonstrated in Sec.~\ref{sec:einstein}, the field equations obtained from the compactification of five-dimensional Einstein-Chern-Simons gravity are structurally identical to the Einstein field equations with a cosmological constant. This structural identity has an important consequence: the observational successes of $\Lambda$CDM cosmology are preserved within this framework, even as its growing tensions motivate the search for a deeper origin of $\Lambda$. The Friedmann equations governing the expansion of the universe, the growth of perturbations, and the formation of large-scale structure remain unchanged. What changes is solely the interpretation and origin of $\Lambda$.

In standard $\Lambda$CDM, the cosmological constant is an unexplained parameter. In the present framework, it emerges geometrically from the compactification of Einstein-Chern-Simons gravity, with its magnitude determined by the compactification radius $r_{c}$ and the coupling constants $l$ and $\tilde{h}$ through the relation~\eqref{eq:lambda}. Thus, the $\Lambda$CDM model can be embedded within a higher-dimensional theory that provides a geometric origin for dark energy, without modifying any of its observational predictions.

\subsection{Two Dynamical Regimes of the Cosmological Constant}

Starting from Eq.~\eqref{eq:lambda}, we identify two physically distinct regimes depending on the relative magnitude of the dimensionless product $\varepsilon V$.

\textit{Regime I: weak-field limit} ($\varepsilon V \ll 1$). In this 
regime, the denominator of Eq.~\eqref{eq:lambda} is dominated by the 
unity term, and the cosmological constant reduces to
\begin{equation}
\Lambda \approx \kappa V = \pm \frac{3 \pi l^{2} \tilde{h}}{8 \kappa^2 r_{c}^{3}}.
\label{eq:lambda_weak}
\end{equation}
In this limit, $\Lambda$ is controlled by the ratio $l^{2}\tilde{h}/r_{c}^{3}$, and the sign of the cosmological constant is determined by the sign of $l^{2}\tilde{h}$. In terms of the vacuum energy density, this reads
\begin{equation}
\kappa \rho_{\text{vac}} = \Lambda = \pm \frac{3 \pi l^{2} \tilde{h}}{8 \kappa^2 r_{c}^{3}}.
\label{eq:rho_weak}
\end{equation}
The condition $\varepsilon V \ll 1$ translates, using Eqs.~\eqref{KK} 
and \eqref{eq:definitions}, into
\begin{equation}
\frac{\pi l^{2} \tilde{h}}{2 r_{c}} \ll \kappa,
\end{equation}
which is satisfied when the Chern-Simons corrections are parametrically small compared to the gravitational coupling. In this regime, $\Lambda$ is directly proportional to the product $l^{2}\tilde{h}$, so that the sign and magnitude of the cosmological constant are controlled jointly 
by the Chern-Simons coupling constant and the trace of the compactified field $h^{a}$.

\textit{Regime II: strong-field limit} ($\varepsilon V \gg 1$). When the Chern-Simons corrections dominate, the cosmological constant saturates to
\begin{equation}
\Lambda \approx \frac{\kappa}{\varepsilon} = \frac{3}{4 r_{c}^{2}},
\label{eq:lambda_strong}
\end{equation}
a remarkable result in which $\Lambda$ depends only on the compactification radius $r_{c}$, becoming entirely independent of the coupling constants $l$ and $\tilde{h}$. This saturation behavior follows from an algebraic cancellation: when $\varepsilon V \gg 1$, the Chern-Simons sector appears in both the numerator and denominator of Eq.~\eqref{eq:lambda}, so that $\Lambda$ approaches the limiting value $3/(4r_{c}^{2})$ independently of $l$ and $\tilde{h}$, with the value of $\Lambda$ determined entirely by the compactification radius $r_{c}$ in this regime.

It is instructive to verify the consistency of the Kottler solution with the sign structure of Table~\ref{tab:branches}. The Kottler metric~\eqref{eq:kottler} is obtained by setting $c_1 = 2M > 0$ and $c_2 = -\Lambda < 0$ in Eq.~\eqref{eq:z8}. Substituting $c_2 = -\Lambda$ into Eq.~\eqref{eq:z9}, with $\varepsilon > 0$ and $\Lambda > 0$, one obtains
\begin{equation}
V = \frac{\Lambda}{\kappa - \varepsilon\Lambda}.
\end{equation}
For $\Lambda > 0$ and $V > 0$, one requires $\kappa - \varepsilon\Lambda > 0$, i.e., $\Lambda < \kappa/\varepsilon$. This corresponds precisely to Branch~1 of Table~\ref{tab:branches}, where $V > 0$ and $1 + \varepsilon V > 0$, yielding $0 < \Lambda < \kappa/\varepsilon = 
3/(4r_c^2)$. Thus, the Kottler solution with $\Lambda > 0$ is consistent with Branch~1, and the compactification radius $r_c$ imposes a geometric upper bound on the admissible values of the 
cosmological constant.

Similarly, the Kottler-AdS solution is obtained by setting $c_2 = -\Lambda > 0$ in Eq.~\eqref{eq:z9}, yielding
\begin{equation}
V = \frac{\Lambda}{\kappa - \varepsilon\Lambda} < 0,
\end{equation}
with $1 + \varepsilon V > 0$, which corresponds to Branch~3 of Table~\ref{tab:branches}. Since $\Lambda < 0$ implies $\kappa - \varepsilon\Lambda = \kappa + \varepsilon|\Lambda| > 0$ 
always, no additional constraint on $\Lambda$ arises from the consistency condition, and the AdS branch is admissible for any $\Lambda < 0$. This AdS branch may be of physical interest in 
light of recent observational studies suggesting the possibility of a negative cosmological constant in the dark sector~\cite{Calderon2021, Adil2023, Menci2024}.

It is worth noting that Branch~2 of Table~\ref{tab:branches}, which requires $V < -1/\varepsilon$, is incompatible with the Kottler solution: substituting $c_2 = -\Lambda$ into Eq.~\eqref{eq:z9} always yields $\varepsilon V > -1$ for any sign of $\Lambda$, since $\kappa > 0$ and $\varepsilon > 0$. Thus, Branch~2 does not correspond to any Kottler-type solution within this framework.

The connection between the branches of Table~\ref{tab:branches} and the two dynamical regimes of 
$\Lambda$ is the following. In Branch~1, the condition $0 < \Lambda < \kappa/\varepsilon = 3/(4r_c^2)$ encompasses both regimes: the weak-field regime ($\varepsilon V \ll 1$), 
where $\Lambda \approx 3\pi l^2\tilde{h}/(8\kappa^2 r_c^3)$ depends explicitly on $l$, $\tilde{h}$, and $r_c$ through Eq.~\eqref{eq:lambda_weak}, and the strong-field regime 
($\varepsilon V \gg 1$), where $\Lambda \to 3/(4r_c^2)$ approaches the asymptotic upper bound independently of $l$ and $\tilde{h}$, as given by Eq.~\eqref{eq:lambda_strong}. In Branch~3, $\Lambda < 0$ is admissible for any value of $\Lambda < 0$ and corresponds exclusively to the weak-field regime ($\varepsilon V \ll 1$), since in the strong-field limit ($\varepsilon V \gg 1$) one always obtains $\Lambda \approx 3/(4r_c^2) > 0$ from Eq.~\eqref{eq:lambda_strong}, independently of the sign of $l^2\tilde{h}$. Thus, a negative cosmological constant is a purely weak-field phenomenon within this framework.

Branch~2 of Table~\ref{tab:branches}, although incompatible with any Kottler-type solution as shown above, admits a natural cosmological interpretation. In this branch, $\Lambda > 
\kappa/\varepsilon = 3/(4r_c^2) \approx \Lambda_{\text{obs}}$, which from the Friedmann equation
\begin{equation}
H^2 = \frac{\Lambda}{3} + \frac{8\pi G}{3}\rho_{\text{tot}},
\end{equation}
implies $H > H_0$, corresponding to epochs earlier than the present. As $V \to -1/\varepsilon$ from below, $\Lambda \to +\infty$, consistent with an early inflationary phase where the cosmological constant was much larger than its present value. As $V \to -\infty$, 
$\Lambda$ decreases asymptotically toward $\kappa/\varepsilon = 3/(4r_c^2)$ from above, suggesting a dynamical transition from Branch~2 toward the present observed value $\Lambda_{\text{obs}} \approx 3/(4r_c^2)$, which lies in Branch~1. Within this framework, 
such a transition would correspond to a dynamical evolution of $V$ from $V < -1/\varepsilon$ toward $V > 0$, driven by the expansion of the universe, and would require passing through the singular point $1 + \varepsilon V = 0$, where $\Lambda$ diverges. This suggests that Branch~2 and Branch~1 are dynamically disconnected, and that the present universe resides permanently in Branch~1.

It is worth noting that the present framework assumes $V$ to be constant, as established by Eq.~\eqref{eq:bianchi}, which leads to a strictly constant $\Lambda$. However, the branch structure of Table~\ref{tab:branches} suggests a richer dynamical picture if $V$ were allowed to evolve with time. In particular, Branch~2 admits a monotonically decreasing $\Lambda(V)$, ranging from $\Lambda \to +\infty$ as $V \to -1/\varepsilon^{-}$ down to $\Lambda \to 
3/(4r_c^2)$ asymptotically as $V \to -\infty$, which could potentially describe a cosmological evolution from an early inflationary phase toward the present observed value $\Lambda_{\text{obs}} \approx 3/(4r_c^2)$. Such a scenario would require a dynamical $V = V(t)$, possibly arising from a time-dependent scalar field $\tilde{\phi}(t)$, and would naturally 
connect Branch~2 with the present universe described by Branch~1. We leave the exploration of this dynamical extension, and its implications for a time-varying cosmological constant, as an 
interesting direction for future work.

\subsection{Estimating the Compactification Radius}

Regime II has a striking observational implication. Inserting the measured value $\Lambda_{\text{obs}} \approx 1.1 \times 10^{-52}\,\text{m}^{-2}$ into Eq.~\eqref{eq:lambda_strong}, we obtain
\begin{equation}
r_{c} = \sqrt{\frac{3}{4\Lambda_{\text{obs}}}} \approx \sqrt{\frac{3}{4 \times 1.1 \times 10^{-52}}} \,\text{m} \approx 8.2 \times 10^{25}\,\text{m},
\label{eq:rc_estimate}
\end{equation}
which is of the order of the Hubble radius,
\begin{equation}
r_{c} \sim H_{0}^{-1} \approx \frac{c}{H_{0}} \approx \frac{3 \times 10^{8}\,\text{m/s}}{2.18 \times 10^{-18}\,\text{s}^{-1}} \approx 1.37 \times 10^{26}\,\text{m}.
\label{eq:rc_hubble}
\end{equation}
More precisely, combining Eq.~\eqref{eq:lambda_strong} with the Friedmann equation for a flat universe dominated by the cosmological constant, $H^{2} = \Lambda c^{2}/3$, one obtains
\begin{equation}
r_{c} = \frac{1}{2} \frac{c}{H_{0}} \sqrt{\frac{1}{\Omega_{\Lambda}}},
\label{eq:rc_friedmann}
\end{equation}
where $\Omega_{\Lambda} \approx 0.685$ is the dark energy density parameter~\cite{Aghanim}. This yields $r_{c} \approx 0.78\, H_{0}^{-1}$, confirming that the compactification radius is comparable to, but distinct from, the Hubble radius. The numerical proximity of $r_{c}$ and $H_{0}^{-1}$ suggests that the size of the fifth dimension and the Hubble horizon may be governed by the same underlying physics.

In contrast, within Regime I, the observational constraint $\Lambda_{\text{obs}} \approx 10^{-52}\,\text{m}^{-2}$ imposes a relation among three free parameters,
\begin{equation}
\frac{l^{2}\tilde{h}}{r_{c}^{3}} \approx \frac{8\kappa \Lambda_{\text{obs}}}{3\pi} \approx 1.9 \times 10^{-53}\,\text{m}^{-3},
\label{eq:constraint_weak}
\end{equation}
leaving a one-parameter family of solutions. For instance, if one imposes Planck-scale values $l \sim l_{\rm Pl} \approx 1.6 \times 10^{-35}\,\text{m}$ and $\tilde{h} \sim 1$, then Eq.~\eqref{eq:constraint_weak} yields $r_{c} \sim 10^{-14}\,\text{m}$, which is sub-nuclear in scale. Conversely, if $r_{c} \sim H_{0}^{-1}$, then $l^{2}\tilde{h} \sim 10^{25}\,\text{m}^{3}$, requiring either $l$ to be macroscopic or $\tilde{h}$ to be very large. This contrast illustrates that only Regime II provides a natural and parameter-independent explanation for the observed value of $\Lambda$.

A further consequence of identifying $\Lambda$ with the compactification radius through Eq.~\eqref{eq:lambda_strong} concerns the thermodynamics of the cosmological horizon of the Kottler solution derived in Sec.~\ref{sec:spherical}. In the limit $M \to 0$, the cosmological horizon radius is determined by $1 - \Lambda r^2/3 = 0$, giving $r_{\rm cosm} = \sqrt{3/\Lambda_{\rm obs}}$. The area of this horizon is then
\begin{equation}
A_{\rm cosm} = 4\pi r_{\rm cosm}^2 = 4\pi \cdot \frac{3}{\Lambda_{\rm obs}} 
= \frac{12\pi}{\Lambda_{\rm obs}}.
\end{equation}
Substituting into the Bekenstein-Hawking formula $S = k_B A/(4\,l_{\rm Pl}^2)$~\cite{GibbonsHawking}, the entropy of the cosmological horizon becomes
\begin{equation}
S_{\rm cosm} = \frac{k_B}{4\,l_{\rm Pl}^2} \cdot \frac{12\pi}{\Lambda_{\rm obs}}
= \frac{3\pi k_B}{\Lambda_{\rm obs}\, l_{\rm Pl}^2} 
= \frac{4\pi k_B r_{c}^{2}}{l_{\rm Pl}^{2}},
\label{eq:entropy_cosm}
\end{equation}
where in the last equality we used $\Lambda_{\rm obs} = 3/(4r_c^2)$ from Eq.~\eqref{eq:lambda_strong}. Using $r_{c} \sim H_{0}^{-1}$, this yields $S_{\rm cosm} \sim 10^{122}\, k_B$, in agreement with the Gibbons-Hawking entropy of the de Sitter horizon. 
The thermodynamics of the Kottler spacetime, including the two-temperature structure associated with the black hole and cosmological horizons, has been studied in detail in~\cite{GibbonsHawking,Bhattacharya}. In this framework, the entropy of the cosmological horizon acquires a geometric interpretation: it is determined by the ratio of the compactification radius to the Planck length.

\subsection{Geometric Reformulation of the Cosmological Constant Problem}

The cosmological constant problem, in its most acute form, asks why the quantum vacuum energy---which by dimensional analysis should be of order $E_{\rm Pl}^{4}/(c\hbar)^{3} \sim 10^{112}\,\text{J/m}^{3}$, corresponding to $\Lambda_{\rm QFT} \sim 10^{69}\,\text{m}^{-2}$---differs from the observed value $\Lambda_{\rm obs} \approx 10^{-52}\,\text{m}^{-2}$ by 121 orders of magnitude. Within the present framework, this question is recast in geometric terms. From Eq.~\eqref{eq:lambda_strong}, the observed value $\Lambda_{\rm obs}$ determines the compactification radius through
\begin{equation}
r_{c} = \sqrt{\frac{3}{4\Lambda_{\rm obs}}} \approx 8 \times 10^{25}\,\text{m}.
\label{eq:rc_problem}
\end{equation}
This geometric restatement does not resolve the fine-tuning, but reformulates it in terms of a length scale: rather than asking why $\Lambda_{\rm obs}/\Lambda_{\rm QFT} \sim 10^{-121}$, one asks why $r_{c} \sim 10^{26}\,\text{m}$. Whether this reformulation offers a more tractable path toward a fundamental explanation remains an open question.

It is instructive to compare this result with the original Kaluza-Klein scenario, where the compactification radius is typically set equal to the Planck length, $r_{c} \sim l_{\rm Pl} 
\approx 10^{-35}\,\text{m}$, so that Kaluza-Klein excitation modes acquire masses beyond the Planck scale and remain experimentally inaccessible~\cite{Overduin}. The present framework inverts this logic: here, $r_{c} \sim H_{0}^{-1}$ is cosmological, suggesting that the extra dimension operates at cosmological rather than microscopic scales. Whether this is observationally viable lies beyond the scope of the present work, though the second Randall-Sundrum model~\cite{randall1} provides a proof of concept that an infinite fifth dimension need not violate known tests of gravity.

\section{Concluding Remarks}
\label{sec:conclusions}

In this work, we have studied the four-dimensional compactification of five-dimensional Einstein-Chern-Simons gravity. Our main results can be summarized as follows:

First, we have shown that the field equations obtained from the compactification coincide with the Einstein field equations with a cosmological constant $\Lambda$ that depends on the 
compactification radius $r_{c}$, the coupling constant $l$, and the trace $\tilde{h}$ of the compactified field. Crucially, the structure of the field equations is identical to that of 
standard General Relativity with $\Lambda$, which means that all known vacuum solutions---including Schwarzschild-de Sitter, Kerr-de Sitter, and FLRW spacetimes---remain valid within this framework. The novelty lies not in the form of the solutions, but in the geometric origin of the cosmological constant.

Second, we have found that the most general static, spherically symmetric solution to these field equations is a Kottler-type (Schwarzschild-de Sitter) black hole, characterized by the line element~\eqref{eq:kottler} with the cosmological constant given by Eq.~\eqref{eq:lambda}.

Third, and most significantly, this framework offers a new perspective on the cosmological constant problem---one of the most pressing challenges in modern physics. Observational
evidence from Type Ia supernovae, the cosmic microwave background, and baryon acoustic oscillations indicates that approximately 68\% of the energy content of the universe
consists of dark energy, well described by $\Lambda \approx 10^{-52}\,\text{m}^{-2}$. Yet quantum field theory, using the Planck scale as a natural cutoff, predicts a vacuum energy
density corresponding to $\Lambda \sim 10^{69}\,\text{m}^{-2}$---some 121 orders of magnitude larger than the observed value. Within the standard $\Lambda$CDM model, the cosmological
constant must be introduced as a fine-tuned free parameter without any fundamental explanation for its magnitude. Moreover, the model faces growing observational tensions---including a statistically significant discrepancy in the determination of the Hubble constant $H_0$, hints of dynamical dark energy with equation of state $w \neq -1$, and a mild but persistent discrepancy in the amplitude of matter fluctuations---that suggest $\Lambda$CDM, while an excellent approximation to reality, may be incomplete.

The mechanism presented here provides a conceptually different approach: $\Lambda$ is not a free parameter but emerges naturally from the compactification of Einstein-Chern-Simons
gravity, with its value determined by the relationship $\Lambda \propto l^{2}\tilde{h}/r_{c}^{3}$, as shown in Eq.~\eqref{eq:lambda}. This scaling implies that, for fixed values of the coupling constants $l$ and $\tilde{h}$, a large compactification radius naturally produces a small cosmological constant. Conversely, the observed value of $\Lambda$ constrains the combination $l^{2}\tilde{h}/r_{c}^{3}$. If the coupling constants are of order unity in natural units, this constraint implies a compactification radius $r_{c}$ of the order of the Hubble radius, $r_{c} \sim H_{0}^{-1} \sim 10^{26}\,\text{m}$. While the individual values of $l$, $\tilde{h}$, and $r_{c}$ remain undetermined at this level of analysis, this scaling behavior suggests a potential connection between the size of the extra dimension and the large-scale structure of the observable universe. It is worth noting that the present framework, with $V$ and $r_c$ constant, preserves the full predictive structure of $\Lambda$CDM without resolving its observational tensions directly; addressing these would require a dynamical extension in which $r_c = r_c(t)$, a natural direction for future work.

Fourth, within the cosmological analysis we identified two distinct dynamical regimes. In the weak-field regime ($\varepsilon V \ll 1$), the cosmological constant takes the form $\Lambda \propto l^{2}\tilde{h}/r_{c}^{3}$, whose sign depends on $l^{2}\tilde{h}$ and requires fine-tuning to reproduce $\Lambda_{\rm obs}$. In the strong-field regime ($\varepsilon V \gg 1$), the dependence on $l$ and $\tilde{h}$ drops out through algebraic cancellation and one obtains $\Lambda \approx 3/(4r_{c}^{2})$, independently of the Chern-Simons coupling. This second regime is particularly significant: it naturally yields $\Lambda_{\rm obs}$ for $r_{c} \approx 8.2 \times 10^{25}\,\text{m} \approx 0.78\,H_{0}^{-1}$, a value consistent with the observed Hubble radius, without any fine-tuning of $l$ or $\tilde{h}$. 

Fifth, a further consequence of the strong-field regime concerns the thermodynamics of the cosmological horizon of the Kottler solution. Using $\Lambda_{\rm obs} = 3/(4r_c^2)$, the Bekenstein-Hawking entropy of the cosmological horizon takes the form $S_{\rm cosm} = 3\pi k_B/(\Lambda_{\rm obs}\,l_{\rm Pl}^2) = 4\pi k_B r_c^2/l_{\rm Pl}^2$, which yields $S_{\rm cosm} \sim 10^{122}\,k_B$ for $r_c \sim H_0^{-1}$, in agreement with the Gibbons-Hawking result~\cite{GibbonsHawking,Bhattacharya}. This provides a geometric interpretation of the horizon entropy in terms of the compactification radius.

It is important to emphasize that this mechanism does not solve the cosmological constant problem in the fundamental sense, since one must still explain why $r_{c}$ takes this particular value. However, it reframes the problem geometrically: rather than asking why $\Lambda$ is so small, one asks why the compactification radius is so large. This geometric reformulation, consistent with the findings of Randall and Sundrum~\cite{randall1}, leads to the intriguing possibility that we could consistently coexist with a large (or even infinite) fifth dimension without violating established gravitational tests.

\section*{Acknowledgments}

M.C.\ acknowledges support from the Direcci\'{o}n de Investigaci\'{o}n y Creaci\'{o}n Art\'{\i}stica de la Universidad del B\'{\i}o-B\'{\i}o through grants No.\ RE2320220 and GI2310339. S.L.\ acknowledges support from FONDECYT grant No.\ 1250969 from the Government of Chile. C.R.\ was supported by ANID fellowship No.\ 2219610 from the Government of Chile and by Universidad de Concepci\'{o}n, Chile. P.S.\ acknowledges support from Grant UNAP VRII No.\ 091/25, Iquique, Chile.

\end{document}